\documentclass[aps,pra,twocolumn,groupedaddress]{revtex4-1}
\usepackage{graphicx}
\usepackage{amsmath}
\usepackage{dcolumn}	
\usepackage{bm}			
\usepackage{amsfonts}
\usepackage{xspace}
\usepackage{color}
\usepackage{epstopdf}
\usepackage{multirow}

\begin{document}
\newcommand{\sss}[1]{{\color{red}#1}}
\newcommand{\KandP}[1]{{\color{red}(#1)}}
\title{Quantum tunneling dynamics of an interacting Bose-Einstein condensate through a Gaussian barrier}

\author{P.~Manju$^{1}$}
\author{K.~S.~Hardman$^{1}$}
\author{M.~A.~Sooriyabandara$^{1}$}
\author{P.~B.~Wigley$^{1}$}
\author{J.~D.~Close$^{1}$}
\author{N.~P.~Robins$^{1}$}
\author{M.~R.~Hush$^{2}$}
\author{S.~S.~Szigeti$^{1}$}
\affiliation{$^{1}$Atomlaser and Quantum Sensors Group, Department of Quantum Science, Research School of Physics and Engineering, The Australian National University, Acton, 2601, Australia\\$^{2}$School of Engineering and Information Technology,University of New South Wales at the Australian Defence Force Academy, Canberra, 2600, Australia}
\date{\today}

\begin{abstract}
    The transmission of an interacting Bose-Einstein condensate incident on a repulsive Gaussian barrier is investigated through numerical simulation. The dynamics associated with interatomic interactions are studied across a broad parameter range not previously explored. Effective 1D Gross-Pitaevskii equation (GPE) simulations are compared to classical Boltzmann-Vlasov equation (BVE) simulations in order to isolate purely coherent matterwave effects. 
    Quantum tunneling is then defined as the portion of the GPE transmission not described by the classical BVE. An exponential dependence of transmission on barrier height is observed in the classical simulation, suggesting that observing such an exponential dependence is \emph{not} a sufficient condition for quantum tunneling. Furthermore, the transmission is found to be predominately described by classical effects, although interatomic interactions are shown to modify the magnitude of the quantum tunneling. Interactions are also seen to affect the amount of classical transmission, producing transmission in regions where the non-interacting equivalent has none. This theoretical investigation clarifies the contribution quantum tunneling makes to overall transmission in many-particle interacting systems, potentially informing future tunneling experiments with ultracold atoms.
    
\end{abstract}

\maketitle

\section{introduction}
Quantum tunneling of a wave packet through a potential barrier is a fundamental quantum mechanical problem that has been extensively studied for decades \cite{TunnelingHartman1962, TunnelingJosephson1962, BarrierPenetrationCohen1965, MacroQTunnelingJosephsonJunction1985, TrunkatedGaussianBarrier1988,Q.GaussianWellsBarriers,GaussianBarrierWronskian}. Beyond its fundamental interest, quantum tunneling is crucial to  technological applications such as the tunnel diode \cite{TunnelDiode1960}, the scanning tunneling microscope \cite{ScanningTunnelingMicroscopy2000}, and SQUIDs \cite{ MacroQTunnelingSQUIDExperimentalPRL2002, ExperimentalQTunneling_ScReports2013, QInterferenceInMeoscopicSCLoopSQUID1993, JosephsonHeatInterferometerSQUID2012Nature}. 
Recent experimental progress in ultracold atomic physics has provided a new, flexible platform in which to explore this phenomenon. These systems are isolated from the environment and offer a high degree of control through a combination of magnetic, optical, and rf fields. Additionally, the high phase-space density of Bose-Einstein condensates (BECs) allow for large interatomic interactions, which can be precisely tuned through a Feshbach resonance. This enables investigations into many-body effects with both attractive and repulsive interatomic interactions. This exquisite control makes BECs ideally suited to detailed studies of quantum tunneling in a wide parameter regime.
 \begin{figure}[t!]
	\includegraphics[width=1\linewidth]{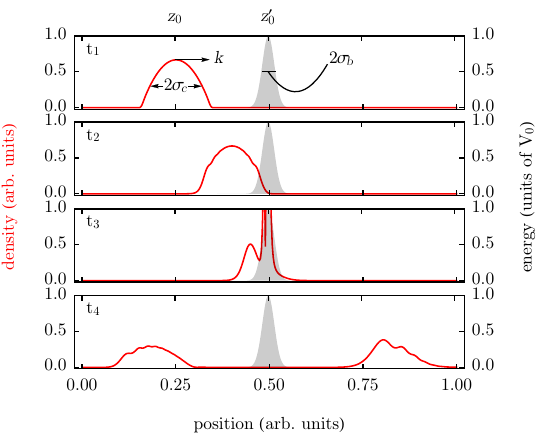}
	\caption{ Schematic representation of a 1D BEC interacting with a Gaussian potential barrier at four snapshots in time. Here the shaded region shows the Gaussian barrier with the red curves representing the density profile of the BEC. The BEC begins on the left side of the barrier ($t_1$), propagates towards it ($t_2$), hits the barrier ($t_3$), and splits into transmitted and reflected components ($t_4$). Key parameters are shown including the initial momentum kick given to the cloud, $k$, the RMS-width of the cloud, $\sigma_c$, the width of the barrier, $\sigma_b$, and the height of the barrier, $V_0$.}
	\label{schematicRepresentation}
\end{figure}

Matterwave transmission and/or reflection has been investigated for ultracold atoms coherently interacting with optical lattices~\cite{SolitonLatticeSpatialGapCarusotto,BolchLikeReflectionSpatialGapSantos,TunnelBarriersUsingSpatialGapsDOdelin}, double-well potentials~\cite{BECJosephsonJnOberthaler2006}, and silicon surfaces \cite{Pasquini:2004, QReflectionKetterlePRL}, and also under driving and dissipative processes~\cite{DissipationTunnelingHerwigOttPRL2016}. Previous theoretical studies into the transmission properties of condensates with interatomic interactions through potential barriers have investigated the emergence of blips~\cite{ElongatedBECTunneling}, the lifetime and stability of quasibound states in a potential well~\cite{BECtunnelingPotentialWell}, and the transmission time~\cite{BECTunnelingTimeRectangular}. 

The transmission properties of a bright soliton through a barrier have also been studied theoretically and experimentally, with the effect of interactions on the transmission coefficient demonstrated~\cite{BrightSolitonsTunneling,SolitonInterferometerTunneling2013,marchant_quantum_2016}. An increase in the transmission rate with atom number has been shown using numerical simulations \cite{BECPulsedTunneling} and demonstrated experimentally \cite{BECInteractionAssistedTunnelingExperimental}, while the dependence of transmission on barrier height has also been experimentally verified~\cite{GuidedAtomLaserAspect2007}. 

Although control of tunneling through manipulation of the atom number is experimentally achievable, the scattering length and potential barrier height provide a simpler pathway to achieving this goal. This paper explores these controls across a broad parameter space not previously studied, detailing the various associated transmission dynamics. 

Furthermore, since transmission in quantum systems is not exclusively a quantum effect, the magnitude of quantum tunneling is difficult to robustly quantify. 
This paper develops a theoretical procedure for isolating quantum tunneling from overall transmission in a many-particle quantum system, and applies this to the simple case of a one-dimensional (1D) BEC interacting with a Gaussian potential barrier.
This procedure compares simulations of the 1D Gross-Pitaevskii equation (GPE), a mean-field model that fully captures the relevant quantum coherent effects, with a Boltzmann-Vlasov equation (BVE), which instead models the atoms as classical hard spheres that interact via a mean-field interatomic potential.
Quantum tunneling is then defined as the difference in transmission between these two models. The system is numerically investigated across a broad parameter space including interatomic interactions that range from fully repulsive through non-interacting to fully attractive. The transmission dynamics associated with changing interactions are shown to be dominated by classical effects in all parameter regimes, however the amplitude of quantum tunneling is also affected. 

\section{Model System and Parameters}
The theoretical system is modelled according to experimentally-realizable parameters \cite{kuhn_bose-condensed_2014,MI_Patrick_2017,Imaging_Paul2016}, briefly described below. A BEC of $10^5$ $^{85}$Rb atoms is initially prepared in a cylindrically-symmetric harmonic trapping potential with radial and axial trapping frequencies $\omega_{\perp}=2\pi \times 70$~Hz and $\omega_{z}=2\pi \times10$~Hz, respectively. This realizes a cigar-shaped condensate where the radial degrees of freedom do not contribute significantly to the dynamics, allowing treatment with a quasi-1D model. Explicitly, the axial trapping potential is
 \begin{align}
	V_\text{trap}(z)=\frac{1}{2}m\omega_{z}^2(z-z_{0})^2, \label{initial_V_HO}
 \end{align}
where $m$ is the mass of a $^{85}$Rb atom. We set $z_{0}=-50 l_z$, where $l_{z}=\sqrt{{\hbar}/{m\omega_{z}}}$. The $s$-wave scattering length of the atoms can be tuned using a Feshbach resonance \cite{SolitonGordonPRL2014, ErratumSolitonGordonPRL, MI_Patrick_2017}. The initial scattering length is set to $5a_{0}$ (where $a_{0}$ is the Bohr radius), giving a Thomas-Fermi density profile with RMS width $\sigma_c=4.7 l_z$. This forms the initial condition for all numerical simulations.
 
 At time $t=0$ the axial trapping potential, $V_\text{trap}(z)$, is extinguished and the scattering length ($a_s$) is quenched from 5$a_{0}$ to a value between $-0.5 a_0$ and $+1 a_{0}$. Simultaneously, the BEC is given a momentum kick along the $z$-direction of $p_0 = 20 \hbar l_z^{-1}$ ($0.73\hbar k_0$, where $k_0$ is the wavenumber of the desired Rb transition). This can be achieved experimentally using Bragg transitions \cite{Ernst:2009,Altin:2013}. This ensures that the condensate has a kinetic energy ${E\sim200 \hbar \omega_z}$ much greater than its initial interatomic interaction potential energy of $\sim{15.9} \hbar \omega_z$. A repulsive Gaussian potential barrier is introduced with potential described by
 \begin{align}
 	V_b(z)	&= V_{0}e^{-(z-z_0')^2/ \sigma_b^2}, \label{potential_barrier}
 \end{align}
where $V_0$ and $\sigma_b$ parametrize the barrier height and width, respectively. This can be created experimentally using a blue detuned laser beam.

The position of the barrier, $z_0'$, is selected such that the initial atomic wavefunction is unperturbed (to machine precision) through the introduction of the barrier.
This condition is achieved by positioning the Gaussian tail of the barrier (at $z = z_0'- 3\sigma_b$) a distance $15 l_z$ from the $3\sigma$-width of the cloud, \emph{i.e.} $z_0' = z_0 + 3(\sigma_b + \sigma_c) + 15 l_z$.
Figure~\ref{schematicRepresentation} gives a schematic representation of the resulting wavefunction evolution. In all simulations, the barrier height, $V_0$, is chosen between $180 \hbar \omega_z$ and $220 \hbar \omega_z$ and the barrier width, $\sigma_b$, is between $0.1\sigma_c$ to $10\sigma_c$. This paper explores the variation in the transmission coefficient as a function of quenched scattering length and various barrier parameters, while keeping the initial condition of the BEC and momentum kick constant.

\section{Methods}\label{Methods}
In order to separate the effect of quantum tunneling from the overall transmission, numerical simulations from two different theoretical approaches are investigated. 
Firstly, the GPE provides an excellent description of the bulk properties of the condensate in the zero-temperature limit and includes relevant quantum effects such as matterwave interference. It therefore provides the total transmission that is experimentally observable.
Secondly, the BVE gives a classical representation of the particle dynamics that includes the interatomic interactions via a mean-field potential while neglecting matterwave effects. The comparison of these two simulations allows one to isolate the purely quantum mechanical behavior. 

\subsection{Gross-Pitaevskii Equation Simulation}
The interatomic dynamics of a BEC in a quasi-1D geometry is described by the 1D GPE
\small
\begin{align}
    i \hbar \frac{\partial \Psi(z ,t)}{\partial t} = &\bigg[ -\frac{\hbar^2}{2 m}\frac{\partial^2}{\partial z^2} + V_\text{ext}(z,t) + g_\text{1D}|\Psi(z,t)|^2 \bigg]\Psi(z,t), \label{GPEEquation1D}
\end{align}
\normalsize
where $V_\text{ext}(z,t)$ is the external axial potential (initially a harmonic potential, then subsequently a Gaussian barrier), $\Psi(z,t)$ is the macroscopic wavefunction (or order parameter) normalized to the total particle number, $N = \int dz |\Psi(z,t)|^2$ with density $\rho(z,t)=|\Psi(z,t)|^2$, and the 1D interaction strength 
\begin{align}
	g_\text{1D}	&=\frac{2^{5/2}}{3N}\frac{1}{\sqrt{m\omega_\text{z}^2}}\left[\frac{15Ng_\text{3D}\omega_\perp^2\omega_z\left(\frac{m}{2}\right)^{3/2}}{8\pi}\right]^{3/5},
\end{align}
where $g_{3D} = 4 \pi \hbar^2 a_s / m$ is the 3D interaction strength produced by a scattering length $a_s$. There are a variety of approaches to deriving an effective 1D GPE from the full 3D GPE \cite{Effective1DGPEPRL98,Effective1DGPEPRA2002,DimensionalDeduction2015}. 
In this paper, a fixed Thomas-Fermi profile is assumed in the radial direction, and the dimensional reduction is performed by equating the chemical potential of the effective 1D GPE to the chemical potential of the full 3D theory. See Appendix~\ref{sec_appendix} for further details.

Equation~(\ref{GPEEquation1D}) is solved using a split-step Fourier and fourth-order Runge-Kutta method. The initial condition for each GPE simulation is
\begin{align}
  \psi_{0}'(z)=\psi_{0}(z)e^{ikz} \label{GPE_IC}
\end{align}
where $\hbar k$ is the magnitude of the momentum kick and $\psi_0(z)$ is the groundstate wavefunction for the initially trapped BEC [\emph{c.f.} Eq.~(\ref{initial_V_HO})] with scattering length $5 a_0$. $\psi_{0}(z)$ is obtained numerically using an imaginary-time propagation method~\cite{ComputeBECGroundStateImaginaryTime,SolveScrodingerEqImaginaryTime}.

\subsection{Boltzmann-Vlasov Equation Simulation}\label{subsection:BVE}
In order to isolate the classical component of the transmission, a classical analog to the GPE is required. The Boltzmann-Vlasov equation (BVE) provides this analog by describing the dynamics of an atom subject to the collective interactions created by a large number of other like-particles without the need for a wavefunction~\cite{MeanFieldLimitVlasovBoltzmann,MeanFieldWithCollisionPRA}. Specifically, the BVE analog to the 1D GPE Eq.~(\ref{GPEEquation1D}) is
\begin{align}
	\frac{\partial\mathcal{P}(z,p,t)}{\partial t} + \frac{p}{m}\frac{\partial\mathcal{P}(z,p,t)}{\partial z} - \frac{\partial  (V_b+V_m) }{\partial z} \frac{\partial \mathcal{P}(z,p,t)}{\partial p} = 0, \label{BV_equation}
\end{align}
where $\mathcal{P}(z,p,t)$ is the phase-space distribution for the atoms, $V_b$ is the barrier potential Eq.~(\ref{potential_barrier}), and $V_m$ is the interatomic potential
\begin{align}
	V_m(z,t) = N g_\text{1D}  \int dp \, \mathcal{P}(z,p,t).
\label{meanfield}
\end{align}
The phase-space distribution is normalized to $\int dz dp \,  \mathcal{P}(z,p,t) = N$ with the marginals $\int dp \mathcal{P}(z,p,t)$ and $\int dz \, \mathcal{P}(z,p,t)$ providing the position and momentum-space densities of the atomic cloud, respectively.

Formally, the mean-field treatment of the interatomic interactions via Eq.~(\ref{meanfield}) follows from a quantum description of collisions that includes the wave-like nature of particles during the scattering process~\cite{TransportPropertiesLhuillierLaloe1982}. This density-dependent potential is therefore the direct semiclassical analog of the nonlinear term in the GPE, Eq.~(\ref{GPEEquation1D}). The wave-like origins of Eq.~(\ref{meanfield}) are distinct from the global coherent nature of the BEC matterwave dynamics, which are certainly not captured by the BVE. Crucially, since the BVE treats atoms as hard spheres it does not permit the tunneling of a single atom through a barrier.

Equation~(\ref{BV_equation}) is simulated using a Monte-Carlo sampling method, whereby $M = 10^4$ random samples of the initial phase-space density, $\mathcal{P}(z,p,0)$, are selected and evolved. Each Monte-Carlo sample (indexed by $i$) can be interpreted as a single classical particle at position $x_i$ with momentum $p_i$ and dynamics governed by Newton's equations of motion
\begin{subequations}
\label{MC_samples}
\begin{align}
	\frac{dz_i}{dt} 	&= \frac{p_i}{m}, \\
 	\frac{dp_i}{dt} 	&= -\frac{d}{dz}\left[ V_b(z) + V_m(z,t) \right] \big|_{z=z_i}.
\end{align}
\end{subequations}
To enable a fair comparison to the GPE simulations, the initial phase-space distribution $\mathcal{P}(z,p,0)$ is determined from the position and momentum distributions of the 1D GPE initial condition, Eq.~(\ref{GPE_IC}). Explicitly, $\mathcal{P}(z,p,0)$ is chosen such that $\int dp \, \mathcal{P}(z,p,0) = |\psi_0(z)|^2$ and $\int dz \, \mathcal{P}(z,p,0) = |\tilde \psi_0(p)|^2$, where $\tilde \psi_0(p) \equiv \int dp \exp[-i (p - \hbar k) z / \hbar] \psi_0(z) / \sqrt{2 \pi \hbar}$. 

The interatomic potential, Eq.~(\ref{meanfield}), requires the position-space density $\rho(z,t) = \mathcal{P}(z,p,t)$. At each time step,  $\rho(z,t)$ is estimated from all $M$ samples using a kernel density estimation technique \cite{KernelEstimationPhysics}. This data smoothing method is commonly used to infer populations from a finite data sample. It has also been used in grid-free simulations of the GPE, which is of particular relevance here~\cite{Mocz:2015, Jiang:2018}. The kernel density estimator for $M$ samples is given by 
\begin{align}
	f_\nu(z)	&= \frac{1}{M}\sum_{i=1}^{M} K_\nu(z-z_i),
\end{align}
where the kernel, $K_\nu$, is a non-negative function and $\nu>0$ is a smoothing parameter. A Gaussian kernel function
\begin{align}
 K_\nu(z) =\frac{1}{\sqrt{2 \pi \nu^2}} e^{-\frac{z^2}{2 \nu^2}}
\end{align}
 is chosen, providing an estimate of $\rho(z,t)$ that is smooth and well-behaved in the tails. For Gaussian kernels, the optimal choice of smoothing parameter is
\begin{align}
	\nu = \Bigg(\frac{4\sigma_z^5}{3M}\Bigg)^\frac{1}{5},
\end{align}
where $\sigma_z$ is the standard deviation of samples in position space at the initial time \cite{KernelEstimationPhysics}. This provides the following estimate for the interatomic potential:
\begin{align}
	V_m(z,t) 	&= N g_\text{1D} f_\nu(z) = \frac{N g_{1D}}{M} \sum_{i=1}^{M} \frac{e^{-\frac{(z-z_i)^2}{2 \nu^2}}}{\sqrt{2 \pi \nu^2}}. 
	\label{V_m_estimate}
\end{align}
This interatomic potential couples the $M$ Monte-Carlo samples such that Eq.~(\ref{MC_samples}) produces a set of $M$ coupled ordinary differential equations that must be solved simultaneously.

This Monte-Carlo procedure for solving the BVE, Eq.~(\ref{BV_equation}), was validated by checking that the breathing (i.e. monopole) mode frequency of a 1D harmonically confined BEC with $a_s = 5a_0$ matched the analytic prediction given in Ref.~\cite{Guery-Odelin:2002}. This mode was excited by inducing a small quench of the trapping frequency, as in Ref.~\cite{Bouchoule:2016}. In particular, this specific validation suggests that Eq.~(\ref{V_m_estimate}) provides a sufficiently accurate estimate of the interatomic potential.

\subsection{Definition of Quantum Tunneling}\label{QTunnelingDefinition}
For both of the numerical approaches, the transmitted and reflected regions are defined as $z>z_\text{T}=(z_0'+2\sigma_\text{b})$ and $z<z_\text{R}=(z_0'-2\sigma_\text{b})$, with the number of atoms in each regime defined as
\begin{subequations}
\begin{align}
    N_\text{T}  &= \int_{z_T}^{\infty} dz \, \rho(z,t_\text{end}), \\
    N_\text{R}  &= \int_{-\infty}^{z_\text{R}} dz \, \rho(z,t_\text{end}),
\end{align}
\end{subequations}
respectively. The stopping time, $t_\text{end}$, for the simulation is chosen such that the transmitted and reflected clouds are well separated from the barrier in position space, and both $N_T$ and $N_R$ have reached asymptotic values. The transmission coefficient is then defined as
\begin{align}
    T= \frac{N_\text{T}}{N-N_\text{lost}},
\end{align}
where $N_\text{lost}$ refers to atoms not included in either region, typically those for which the velocity goes to zero after interaction with the barrier. The typical magnitude of $N_\text{lost}$ for current simulations is negligible, on the order of $10^{-6}$. In contrast, $N_\text{loss}$ can be significant for transmission through non-Gaussian barriers such as a square potential barrier and for the transmission of bright solitons through attractive potential barriers \cite{BrightSolitonTunnelingAttractiveBarrierDOdelin2016,marchant_quantum_2016}.

Due to the stochastic nature of the BVE simulation, 20 realizations for each parameter are performed, with the mean transmission coefficient calculated.

The quantum tunneling, $\Delta T$, is then defined as the difference between the transmission coefficient obtained from the GPE simulation and that of the corresponding BVE:
\begin{align}
\Delta T \equiv T_\text{GPE} - T_\text{BVE}.
\label{eq:QuantumTunnelingDefinition}
\end{align}

\section{Results}

\subsection{Total Transmission Using GPE}
\begin{figure}[t]
	\includegraphics[width=1\linewidth]{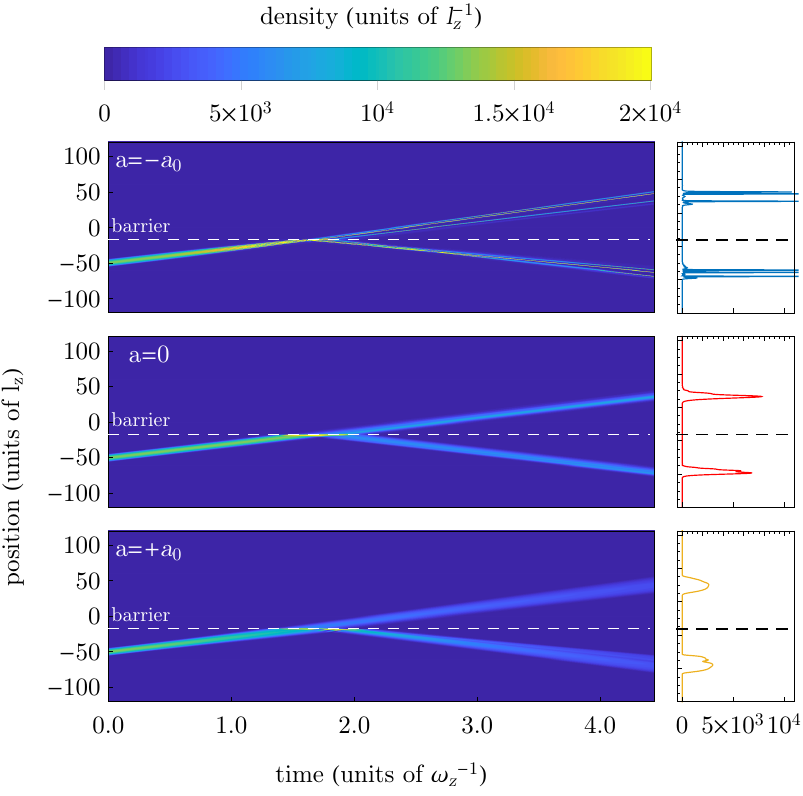}
	\caption{The density profile in position space during a propagation period of $t = 2 \omega_z^{-1}$ for $a_s = \{-1.0,\,0.0,\,1.0\}a_0$ with a barrier height of $V_{0}/E=1$ and $\sigma_b /l_z = 1$. The dotted line represents the central position of the barrier. The density of the cloud at the final time is shown on the right side of each plot. During free propagation, the density distribution expands for $a_s=+1a_0$, contracts for $a_s=-1a_0$, and remains unchanged for $a_s=0$. See the Supplemental Material for a video depicting the real-time evolution of the density.}
	\label{densityPosition}
\end{figure}

First, the GPE dynamics of a condensate through a Gaussian barrier with $\sigma_b / l_z=1$ and $V_0/E=1$ are investigated. Figure~\ref{densityPosition} illustrates the change in density profile during propagation for various scattering lengths. 
As predicted, the non-interacting case where $a_s=0$ is non-focusing during evolution, while attractive ($a_s<0$) and repulsive ($a_s>0$) interactions display focusing and dispersive behavior, respectively \cite{BECNonlinearDynamicsTunneling,BECtunnelingPotentialWell}. The final density profiles illustrated in Fig.~\ref{densityPosition} show the presence of more than one
density peak within the transmitted and reflected components and are more pronounced for the attractive scattering length simulation. This behavior has previously been observed in Ref.~\cite{BECPulsedTunneling}.

\subsubsection{Varying Barrier Width}\label{GPETvsSigma}
\begin{figure}[t]
 	\includegraphics[width=1\linewidth]{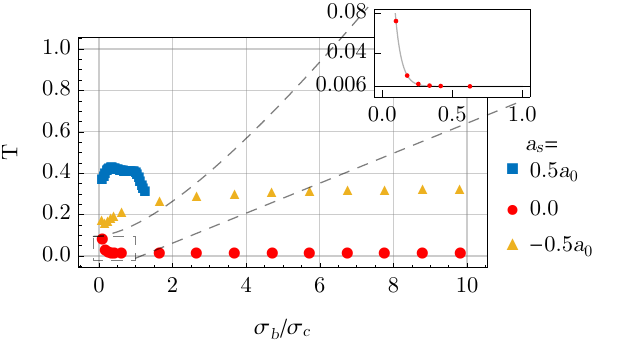}
 	\caption{The variation of the transmission, $T$, with changing potential barrier width, $\sigma_b$, for $V_0/E = 1.1$ and scattering lengths $a_s=\{-0.5,\,0.0,\,0.5\}a_0$, shown in yellow triangles, red dots, and blue squares respectively. Inset shows the non-interacting case with an exponential fit of the form $A\exp\left(-\lambda x\right)+B$ shown as the gray curve. This fit has an r-squared of 0.9998 for parameters $\{A,\,\lambda,\,B\}=\{0.58,\,22.1,\,0.0061\}$. The fit is seen to asymptote to a non-zero value, indicating the presence of a classical contribution to the transmission.}
 	\label{fig:GPEBarrierWidthSims}
\end{figure}

It is well established that the transmission in a non-interacting system displays an exponential dependence on the barrier width in the limit that $V_0>E$. In this regime, the average kinetic energy of the cloud is lower than the barrier, ensuring the bulk of the cloud is unable to pass through and quantum effects dominate. This behavior is confirmed through GPE simulations with $a_s=0$ and a constant potential height $V_0/E=1.1$, as illustrated by the red points in Fig.~\ref{fig:GPEBarrierWidthSims}. Generally, it is expected that the transmission approaches zero as the barrier width goes to infinity. However, these simulations show a non-zero offset, as shown in the inset of Fig.~\ref{fig:GPEBarrierWidthSims}. This offset suggests that the transmission for large barrier widths is dominated by classical dynamics instead of quantum tunneling, implying that transmission is \emph{not} strictly equivalent to quantum tunneling even in the non-interacting regime. This is quantified in Sec.~\ref{subsection:IsolatingQuantumTunneling} by comparing the full dynamics described by GPE simulations to that of classical simulations provided by the BVE. 
 
The introduction of interatomic interactions modifies the simple exponential dependence of transmission on barrier width. This is seen in Fig.~\ref{fig:GPEBarrierWidthSims} where the transmission for $a_s=\{-0.5,0.5\}a_0$ is shown for varying barrier widths. In both attractive and repulsive scattering length regimes, the transmission does not display an exponential dependence on barrier width, instead displaying flat or increasing transmission. Due to the complexity of this behavior, it is not clear whether these effects are due to quantum tunneling or simply classical transmission. The high densities that occur for large barrier widths in the attractive case limit the range of barrier widths where the simulations remain valid.  As with the non-interacting case, these parameters are investigated in Sec.~\ref{subsection:IsolatingQuantumTunneling} using the classical model provided by the BVE. 

Although an analysis of barrier width provides intuitive insight, it is difficult to achieve experimentally where dynamic control of the barrier width requires adaptive optics or spatial light modulators \cite{gauthier_direct_2016}.
By comparison, dynamic control of the barrier height is readily achievable in an experimental setup, motivating further simulations in this parameter space.

\subsubsection{Varying Scattering Length and Barrier Height}\label{GPETvsAsV0}
\begin{figure}[t]
 	\includegraphics[width=1\linewidth]{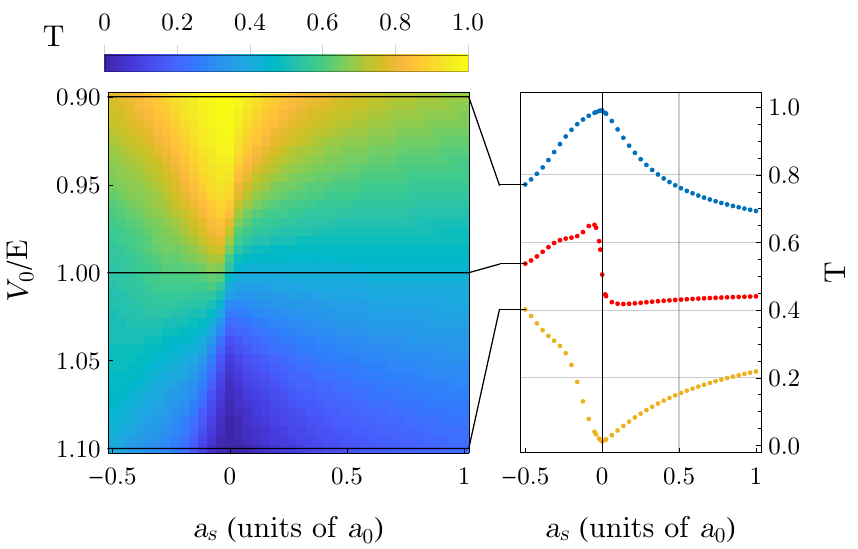}
 	\caption{The variation of transmission, $T$, with barrier height, $V_0$, and scattering length, $a_s$, for constant barrier width $\sigma_b/l_z = 1$. Cross sections taken at $V_0/E=\{0.9,\,1.0,\,1.1\}$ are plotted on the right in blue (upper), red (middle), and yellow (lower) dots}, respectively.
 	\label{SurfacePlot}
 \end{figure}

The dependence of $T$ on scattering length and the ratio of barrier height to the kinetic energy of the cloud is presented in Fig.~\ref{SurfacePlot} for a constant barrier width of $\sigma_b / l_z=1$. For fixed $a_s$, transmission is shown to decrease with increasing barrier height. Cross sections of the surface plot in Fig.~\ref{SurfacePlot} at $V_0/E = \{0.9, 1.0, 1.1\}$ illustrate the general behavior. The result in the regime $E\approx V_0$ and $a_s>0$ is in qualitative agreement with previous theoretical work that explored only this regime~\cite{BECPulsedTunneling}. However, Ref.~\cite{BECPulsedTunneling} suggested that increasing interatomic interactions always enhances the quantum tunneling rate in the quasi-1D regime with a cigar-shaped initial condition. Figure~\ref{SurfacePlot} shows that this is not true in general. Consider the $a_s > 0$ regime, for instance. Although transmission decreases for increasing interatomic interactions when $E>V_0$, it increases with increasing interactions when $E<V_0$.

In general, both attractive and repulsive interatomic interactions nontrivially affect the transmission. Although interatomic interactions modify the overall potential experienced by the atoms, creating an effective potential of the form \cite{BECTunnelingTimeRectangular}
\begin{align}
 V_\text{eff}   &=V_b(z)+g_\text{1D}|\Psi(z,t)|^2, \label{eq_V_eff}
\end{align}
this is not simple to interpret. Na\"ively, this modification increases the effective potential for positive scattering lengths and decreases it for negative scattering lengths, leading one to assume that transmission always increases (decreases) with more attractive (repulsive) interatomic interactions. However, this picture does not account for the dynamical changes in the density due to interatomic interactions. This leads to the transmission's more complicated dependence on $a_s$ and $V_0$, quantified in Fig.~\ref{SurfacePlot}.

Nevertheless, Eq.~(\ref{eq_V_eff}) suggests that some changes in the transmission due to the presence of interatomic interactions arise from energy considerations associated with the size and sign of $g_\text{1D}$ and the dynamics of the density distribution. These considerations are classical and independent of coherent matterwave effects. To quantify separately the effect of interactions on the classical portion of the transmission and the coherent matterwave portion, which is defined as quantum tunneling, the GPE simulations must be compared with the classical BVE simulations.

\subsection{Isolating Quantum Tunneling from Transmission}\label{subsection:IsolatingQuantumTunneling}
GPE simulations alone are unable to isolate the fraction of the transmission associated with coherent matterwave effects. Nevertheless, this can be achieved through comparison with BVE simulations of the classical dynamics. 

\begin{figure*}[t]
	\includegraphics[width=\textwidth]{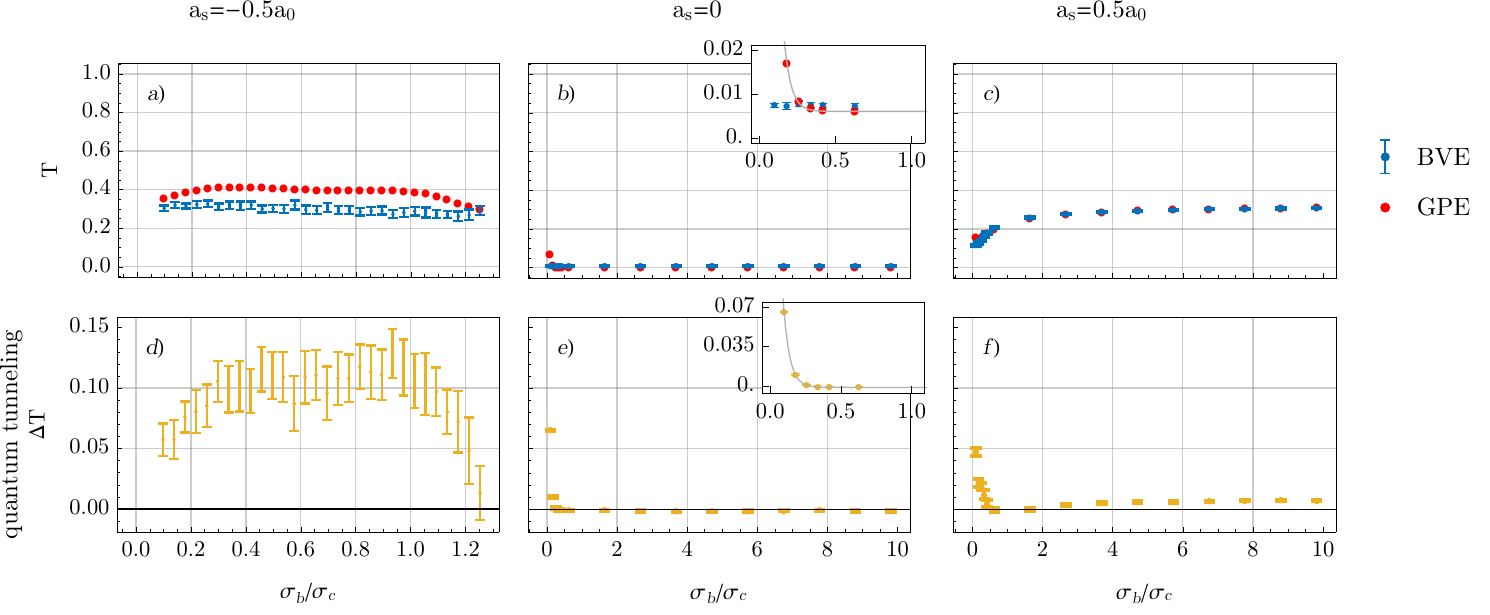}
	\caption {The variation of the transmission coefficient, $T$, with the width of barrier for varying scattering length, $a_s =\{-0.5,\,0.0,\,0.5\}a_0$. The red dot points in (a), (b), and (c) show the transmission calculated from the full GPE simulations, $T_\text{GPE}$, while the blue points with error bars show the transmission computed from the classical BVE simulations, $T_\text{BVE}$. The error bars indicate three times the standard error in the mean as calculated from $20$ realizations for each parameter. (d), (e), and (f) plot the quantum tunneling, defined as the difference between the full GPE and classical BVE simulations [see Eq.~(\ref{eq:QuantumTunnelingDefinition})]. The insets in (b) and (e) show the exponential dependence on barrier width for $a_s=0.0$, with exponential fits of the form $A\exp\left(-\lambda x\right)+B$ shown as the gray curve. The fit in b) results in an r-squared of 0.9998 for parameters $\{A,\,\lambda,\,B\}=\{0.58,\,22.1,\,0.0061\}$, while the fit in e) results in an r-squared of 0.9998 for parameters $\{A,\,\lambda,\,B\}=\{0.58,\,22.1,\,-0.0013\}$. }
	\label{TvsSigma}
\end{figure*}

\subsubsection{Varying Barrier Width}
As in Sec.~\ref{GPETvsSigma}, the effect of changing the barrier width on transmission is simulated, now in a classical system. This is shown as the blue points in Fig.~\ref{TvsSigma} for $a_s=\{-0.5,0.0,0.5\}a_0$ and $V_0/E=1.1$. In the case of a non-interacting system, a similar non-zero value for transmission, even at large barrier widths, is observed, suggesting this effect results from classical effects. This classical transmission arises due to a fraction of the atoms possessing an energy greater than the potential barrier. This fraction is given by the momentum distribution of the cloud.
Figure~\ref{TvsSigma}e) shows the quantum tunneling as defined by Eq.~(\ref{eq:QuantumTunnelingDefinition}) and displays the expected exponential dependence on barrier width. Furthermore, $\Delta T$ approaches zero for increasing barrier width, consistent with models of quantum tunneling.

Similarly, comparing the GPE and BVE simulations for the interacting cases uncovers the portion of the transmission associated with quantum tunneling. However, the dynamics in the interacting regime become significantly more complex. For repulsive ($a_s > 0$) interactions, the quantum tunneling decreases exponentially with increases in barrier width for smaller barrier widths ($\sigma_b<\sigma_c$) and starts increasing for larger widths ($\sigma_b>\sigma_c$) as shown in Fig.~\ref{TvsSigma}d). The attractive interatomic interactions also generate nontrivial quantum tunneling behavior, as shown in Fig.~\ref{TvsSigma}f). Here, dynamics within the barrier provide a major contribution to the transmission and tunneling. For narrow barriers ($\sigma_b<\sigma_c$), there is a chance for multiple reflections and interference inside the barrier. Then the density can be higher and the atoms can be in a regime of strong repulsive or attractive interactions. As the barrier width increases, $\sigma_b>\sigma_c$, the probability of reflections decreases~\cite{SuperluminalTraversalTime}. Ultimately the density, the interatomic interaction energies, and the probability of interference are different for these two barrier width regimes. This causes the transmission to follow a different barrier width dependence in these two regimes.

\begin{figure}
 	\includegraphics[width=1\linewidth]{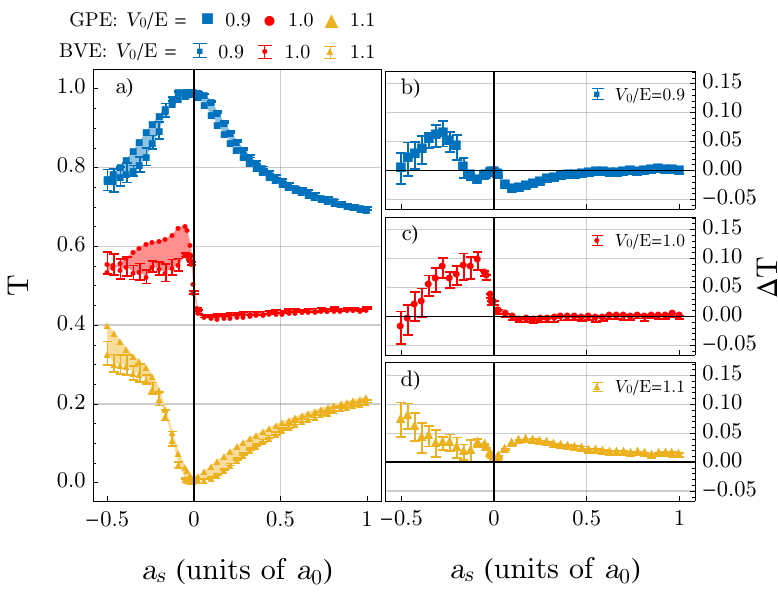}
 	\caption{(a) shows the variation of $T_\text{{GPE}}$ and $T_\text{{BVE}}$ over a range of scattering length, $a_s$, for varying barrier heights, $V_0/E = \{0.9,\,1.0,\,1.1\}$. (b), (c) and (d) show the contribution of quantum tunneling, $\Delta T$, for this same parameter regime. The error bars associated with the BVE simulations indicate three times the standard error in the mean as calculated from $20$ realizations for each parameter.}                     
 	\label{Combined_Classical_GP_Quantum_subplots}
\end{figure}
\subsubsection{Varying Scattering Length} 
Scattering length provides a straightforward experimental method for controlling the chemical potential of the condensate, thereby directly affecting tunneling. This relationship is shown in Fig.~\ref{Combined_Classical_GP_Quantum_subplots} where both GPE and BVE simulation methods are performed for three potential barrier heights $V_0/E=\{0.9,1.0,1.1\}$ and a constant barrier width of $\sigma_b / l_z=1$. The classical simulations display the same qualitative behavior as the total transmission, yet produce distinct quantitative differences. When the cloud has kinetic energy greater than the barrier such that $E>V_0$, transmission in the non-interacting case is high. Interatomic interactions reduce the transmission. As in Ref.~\cite{Damon:2014}, studying the momentum distribution of the cloud during propagation provides insight into this scattering-length dependence of the transmission. In the presence of interatomic interactions, the momentum distribution of the cloud expands during propagation. This results in an increase in the number of atoms with energy less than the barrier and therefore a decrease in the transmission relative to the non-interacting case.

In a similar way, momentum diffusion causes an increase in the transmission when the energy of the cloud is lower than that of the barrier. In this case, the non-interacting limit generates very little transmission, however momentum diffusion from the introduction of interactions allows a portion of the atoms to be at high enough momentum to transmit through the barrier. 

Furthermore, in the presence of interactions, the barrier potential is modified, as discussed in Sec.~\ref{GPETvsAsV0}. This modification results in an enhancement of transmission for attractive interactions and a reduction for repulsive interactions. 

These qualitative effects are present in both the full GPE simulations and those of the classical BVE, suggesting that changes in transmission associated with changing interactions are predominantly classical and not necessarily related to quantum tunneling.

Despite this qualitative agreement between the full GPE simulations and the classical BVE simulations, the degree of quantum tunneling is also affected by interatomic interactions. This is illustrated in Figs.~\ref{Combined_Classical_GP_Quantum_subplots} b), c) and d), where the quantum tunneling, $\Delta T$, is shown for $V_0/E=\{0.9, 1.0, 1.1\}$. The quantum tunneling is calculated as the difference between the GPE and BVE simulations [see Eq.~(\ref{eq:QuantumTunnelingDefinition})] and shown graphically as the shaded region in Fig.~\ref{Combined_Classical_GP_Quantum_subplots}a). It is seen that classical transmission dominates in some regions with no quantum tunneling present. Indeed, certain regions display negative quantum tunneling, implying that matterwave interference effects actually result in a \emph{reduction} of transmission, caused by anti-tunneling~\cite{SpaceMomRepresentationHartmann}.

\begin{figure}[t]
	\includegraphics[width=1\linewidth]{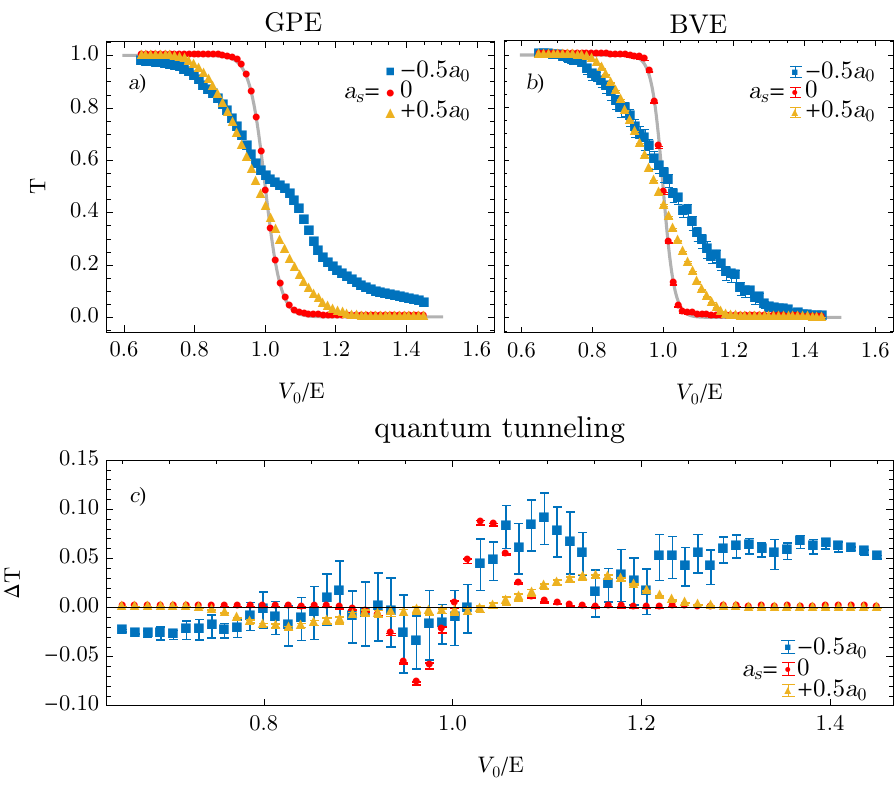}
	\caption {(a) and (b) show the variation of $T_\text{GPE}$ and $T_\text{BVE}$, respectively, for different barrier heights, $V_0/E$, and scattering lengths, $a_s = \{-0.5,\,0.0,\,0.5\}a_0$. (c) shows the contribution of quantum tunneling, $\Delta T $, over this same parameter regime. The error bars are due to the finite sampling error of the BVE simulations and indicate three times the standard error in the mean as calculated from $20$ realizations for each parameter. The quantum tunneling is seen to display a significantly different relationship to barrier height compared to the total transmission, while the classical transmission is in qualitative agreement. This suggests the total transmission is predominantly due to classical effects. The gray curve indicates fits to the non-interacting case for the GPE and BVE simulations using a function of the form $a \tanh\left[b (x - 1)\right]+c$. The GPE fit resulted in an r-squared of $0.999987$ for parameters $\{a,\,b,\,c\}=\{0.500,\,-22.96,\,0.501\}$, while the BVE fit resulted in an r-squared of $0.9999$ for parameters $\{a,\,b,\,c\}=\{0.499,\,31.79,\,0.500\}$.}
	\label{Tr_vs_v0_Classical_1000Samples_GP}
\end{figure} 
	
\subsubsection{Varying Barrier Height}
As with scattering length, the potential barrier height is a readily-accessible experimental control. As such, GPE and BVE simulations are again used to understand the dynamics due to changes in this parameter. Once again, the full GPE simulations are compared to the classical BVE simulations with the results shown in Figs.~\ref{Tr_vs_v0_Classical_1000Samples_GP} a) and b). The exponential dependence of transmission on barrier height, usually considered a sign of quantum tunneling, is also shown to exist for the fully classical transmission. This suggests that observing an exponential dependence on transmission is insufficient to confirm the presence of quantum tunneling. The relationship of quantum tunneling to barrier height is shown in Fig.~\ref{Tr_vs_v0_Classical_1000Samples_GP}c). In the non-interacting regime, quantum tunneling is relatively symmetric around $V_0/E=1$, yet it displays antitunneling behavior for $E>V_0$ and quantum tunneling when $E<V_0$. In the limit of very small and very high potential barriers, quantum tunneling approaches zero in the non-interacting and repulsive regimes. Attractive interactions display quantum tunneling or anti-tunneling in the presence of all the barrier heights investigated here.
As in the previous section, the classical transmission displays a qualitatively-similar relationship to the total transmission, suggesting that the predominant contribution to transmission is from classical effects. In contrast, quantum tunneling displays a vastly different relationship to total transmission.
	
\section{conclusion}
The dynamics of an interacting Bose-Einstein condensate incident on a Gaussian potential barrier has been studied through effective 1D GPE simulations and classical BVE simulations. The comparison of these two simulation methods illustrates the difference between transmission and quantum tunneling. Indeed, an exponential dependence of transmission on barrier height is observed to be present even for classical transmission, suggesting that this property is an insufficient indication of quantum tunneling. Quantum tunneling, defined as the difference between the total transmission and the classical transmission, was investigated, with transmission shown to be predominantly dominated by classical effects for many choices of barrier heights and widths. Coherent matterwave effects even appear to reduce the transmission for certain barrier heights. The tunneling dynamics was studied across a broad parameter space not previously explored.  
Quantum tunneling was seen to be directly controllable through manipulation of the interatomic interactions, a readily achievable experimental control through use of a Feshbach resonance. The simulations used experimentally realizable parameters, which along with non-destructive imaging techniques~\cite{Imaging_Paul2016} suggest that it would be possible to observe tunneling dynamics in real time. These experiments could verify results from the GPE simulations. Additionally, experimental data could be incorporated into semi-empirical classical simulations via the BVE in order to experimentally study quantum tunneling dynamics.

\section{Acknowledgements}
We acknowledge useful discussions with A.~Kordbacheh, P.~J.~Everitt, C.~Freier, S.~Legge, and D.~Pulford. We gratefully acknowledge the support of the NVIDIA Corporation who donated the Tesla K40 GPU used for this research. This research was undertaken with the assistance of resources and services from the National Computational Infrastructure (NCI), which is supported by the Australian Government. The Atomlaser and Quantum Sensors group acknowledge funding from the Australian Research Council projects DP160104965 and DP150100356.

\section{Declarations}
M.~R.~Hush contributed to this work when he was an employee of UNSW Canberra. He now works at Q-CTRL, a private company developing control solutions for quantum computers.

\appendix

\section{Derivation of Effective 1D GPE, Eq.~(\ref{GPEEquation1D})} \label{sec_appendix}
 Consider the 3D GPE describing the macroscopic wavefunction of a BEC:
\begin{align} \label{3D_GPE}
	i \hbar \frac{\partial \Psi(\mathbf{r} ,t)}{\partial t} = &\bigg[ -\frac{\hbar^2}{2 m}\nabla^2 + V_\text{ext}(\mathbf{r},t) + g_\text{3D}|\Psi(\mathbf{r},t)|^2\,, \bigg]\Psi(\mathbf{r},t),
\end{align}
where $V_\text{ext}$ is the external harmonic trapping potential with cylindrical symmetry described by
\begin{align}
    V_\text{ext}=\frac{1}{2}m(\omega_\perp^2 r^2+\omega_z^2z^2)\,,
\end{align}
where $m$ is the atomic mass, and $\omega_\perp$ and $\omega_z$ are the transverse and longitudinal trapping frequencies, respectively. 

For large atom number and repulsive interactions, the interaction energy of the condensate is significantly larger than the kinetic energy. Here the Thomas-Fermi approximation can be applied, which gives
\begin{subequations}
\begin{align}
\mu_{3D}    &=\frac{1}{2}m(\omega_\perp^2 r^2+\omega_z^2z^2)+g_\text{3D}|\Psi(\mathbf{r},t)|^2,\\
|\Psi(\mathbf{r},t)|^2  &=\frac{1}{g_\text{3D}}\left[\mu_\text{3D}-\frac{1}{2}m(\omega_\perp^2 r^2+\omega_z^2z^2)\right],
\end{align}
\end{subequations}
where $\mu_{3D}$ is the chemical potential.
The normalization condition gives
\begin{align}\label{N_3D}
N   &=\int_{0}^{2\pi} d\theta\int_{-z_\text{TF}}^{z_\text{TF}} dz \int_{0}^{r_\text{TF}} dr r |\Psi(\mathbf{r},t)|^2 \notag \\
    &=\frac{8\pi}{15}\frac{\omega_z^2\mu_\text{3D}}{g_\text{3D}\omega_\perp^2}\bigg(\frac{2\mu_\text{3D}}{m\omega_z}\bigg)^{3/2},
\end{align}
where
\begin{subequations}
\begin{align}
    r_\text{TF} &=\sqrt{\frac{2\mu}{m\omega_\perp^2}-\frac{\omega_z^2z_\text{TF}}{\omega_\perp^2}}, \\
    z_\text{TF} &=\sqrt{\frac{2\mu_\text{3D}}{m\omega_z^2}}.
\end{align}
\end{subequations}

Using Eq.~(\ref{N_3D}),
\begin{align}
\mu_\text{3D}=&\left[\frac{15Ng_\text{3D}\omega_\perp^2\omega_z}{8\pi}\big(\frac{m}{2}\big)^{3/2}\right]^{2/5}.
\label{mu3D}
\end{align}
The chemical potential in 1D, $\mu_\text{1D}$, can be similarly obtained by neglecting the kinetic energy in the 1D GPE, Eq.~(\ref{GPEEquation1D}), and it is given by
\begin{align}
\mu_{1D}    &=\frac{1}{2}\left[\frac{3}{2}g_\text{1D}N\sqrt{m\omega_z^2}\right]^{2/3}.
\label{mu1D}
\end{align}
The key step of the dimensional reduction is to choose $g_\text{1D}$ such that the bulk dynamics of the effective 1D GPE well-approximate the true 3D dynamics given by Eq.~(\ref{3D_GPE}). This is provided by setting $\mu_\text{1D} = \mu_\text{3D}$, which matches the energy per particle of the 1D groundstate to the 3D groundstate. Equating Eq.~(\ref{mu3D}) and Eq.~(\ref{mu1D}) and solving for $g_\text{1D}$ gives Eq.~(\ref{GPEEquation1D}).

\bibliographystyle{unsrt}
\bibliography{tunnelingRevised}

\begin{thebibliography}{10}

\bibitem{TunnelingHartman1962}
Thomas~E. Hartman.
\newblock Tunneling of a wave packet.
\newblock {\em Journal of Applied Physics}, 33(12):3427--3433, 1962.

\bibitem{TunnelingJosephson1962}
B.D. Josephson.
\newblock Possible new effects in superconductive tunnelling.
\newblock {\em Physics Letters}, 1(7):251 -- 253, 1962.

\bibitem{BarrierPenetrationCohen1965}
Bernard~L. Cohen.
\newblock A simple treatment of potential barrier penetration.
\newblock {\em American Journal of Physics}, 33(2):97--98, 1965.

\bibitem{MacroQTunnelingJosephsonJunction1985}
Michel~H. Devoret, John~M. Martinis, and John Clarke.
\newblock Measurements of macroscopic quantum tunneling out of the zero-voltage
  state of a current-biased josephson junction.
\newblock {\em Phys. Rev. Lett.}, 55:1908--1911, Oct 1985.

\bibitem{TrunkatedGaussianBarrier1988}
Jeff~D. Chalk.
\newblock A study of barrier penetration in quantum mechanics.
\newblock {\em American Journal of Physics}, 56(1):29--32, 1988.

\bibitem{Q.GaussianWellsBarriers}
Francisco~M. Fern{\'a}ndez.
\newblock Quantum gaussian wells and barriers.
\newblock {\em American Journal of Physics}, 79(7):752--754, 2011.

\bibitem{GaussianBarrierWronskian}
Francisco~M. Fern{\'a}ndez.
\newblock Wronskian method for one-dimensional quantum scattering.
\newblock {\em American Journal of Physics}, 79(8):877--881, 2011.

\bibitem{TunnelDiode1960}
Leo Esaki and Yuriko Miyahara.
\newblock A new device using the tunneling process in narrow p-n junctions.
\newblock {\em Solid-State Electronics}, 1(1):13 -- 21, 1960.

\bibitem{ScanningTunnelingMicroscopy2000}
G.~Binnig and H.~Rohrer.
\newblock Scanning tunneling microscopy.
\newblock {\em IBM Journal of Research and Development}, 44(1.2):279--293, Jan
  2000.

\bibitem{MacroQTunnelingSQUIDExperimentalPRL2002}
Shao-Xiong Li, Yang Yu, Yu~Zhang, Wei Qiu, Siyuan Han, and Zhen Wang.
\newblock Quantitative study of macroscopic quantum tunneling in a dc squid: A
  system with two degrees of freedom.
\newblock {\em Phys. Rev. Lett.}, 89:098301, Aug 2002.

\bibitem{ExperimentalQTunneling_ScReports2013}
Guan-Ru Feng, Yao Lu, Liang Hao, Fei-Hao Zhang, and Gui-Lu Long.
\newblock Experimental simulation of quantum tunneling in small systems.
\newblock {\em Scientific Reports}, 3:2232, August 2013.

\bibitem{QInterferenceInMeoscopicSCLoopSQUID1993}
V.~V. Moshchalkov, L.~Gielen, M.~Dhalle, C.~Van~Haesendonck, and
  Y.~Bruynseraede.
\newblock Quantum interference in a mesoscopic superconducting loop.
\newblock {\em Nature}, 361(6413):617--620, 02 1993.

\bibitem{JosephsonHeatInterferometerSQUID2012Nature}
Francesco Giazotto and Maria~Jose Martinez-Perez.
\newblock The josephson heat interferometer.
\newblock {\em Nature}, 492(7429):401--405, 12 2012.

\bibitem{SolitonLatticeSpatialGapCarusotto}
Iacopo Carusotto, Davide Embriaco, and Giuseppe~C. La~Rocca.
\newblock Nonlinear atom optics and bright-gap-soliton generation in finite
  optical lattices.
\newblock {\em Phys. Rev. A}, 65:053611, May 2002.

\bibitem{BolchLikeReflectionSpatialGapSantos}
Luis Santos and Luis Roso.
\newblock Bloch-like quantum multiple reflections of atoms.
\newblock {\em Phys. Rev. A}, 60:2312--2318, Sep 1999.

\bibitem{TunnelBarriersUsingSpatialGapsDOdelin}
P.~Cheiney, F.~Damon, G.~Condon, B.~Georgeot, and D.~Gu{\'e}ry-Odelin.
\newblock Realization of tunnel barriers for matter waves using spatial gaps.
\newblock {\em EPL (Europhysics Letters)}, 103(5):50006, 2013.

\bibitem{BECJosephsonJnOberthaler2006}
R.~Gati, M.~Albiez, J.~F{\"o}lling, B.~Hemmerling, and M.K. Oberthaler.
\newblock Realization of a single josephson junction for bose--einstein
  condensates.
\newblock {\em Applied Physics B}, 82(2):207--210, Feb 2006.

\bibitem{Pasquini:2004}
T.~A. Pasquini, Y.~Shin, C.~Sanner, M.~Saba, A.~Schirotzek, D.~E. Pritchard,
  and W.~Ketterle.
\newblock Quantum reflection from a solid surface at normal incidence.
\newblock {\em Phys. Rev. Lett.}, 93:223201, Nov 2004.

\bibitem{QReflectionKetterlePRL}
T.~A. Pasquini, M.~Saba, G.-B. Jo, Y.~Shin, W.~Ketterle, D.~E. Pritchard, T.~A.
  Savas, and N.~Mulders.
\newblock Low velocity quantum reflection of bose-einstein condensates.
\newblock {\em Phys. Rev. Lett.}, 97:093201, Aug 2006.

\bibitem{DissipationTunnelingHerwigOttPRL2016}
Ralf Labouvie, Bodhaditya Santra, Simon Heun, and Herwig Ott.
\newblock Bistability in a driven-dissipative superfluid.
\newblock {\em Phys. Rev. Lett.}, 116:235302, Jun 2016.

\bibitem{ElongatedBECTunneling}
G.~Dekel, V.~Farberovich, V.~Fleurov, and A.~Soffer.
\newblock Dynamics of macroscopic tunneling in elongated bose-einstein
  condensates.
\newblock {\em Phys. Rev. A}, 81:063638, Jun 2010.

\bibitem{BECtunnelingPotentialWell}
L~D Carr, M~J Holland, and B~A Malomed.
\newblock Macroscopic quantum tunnelling of bose--einstein condensates in a
  finite potential well.
\newblock {\em Journal of Physics B: Atomic, Molecular and Optical Physics},
  38(17):3217, 2005.

\bibitem{BECTunnelingTimeRectangular}
Zhenglu Duan, Bixuan Fan, Chun-Hua Yuan, Jing Cheng, Shiyao Zhu, and Weiping
  Zhang.
\newblock Quantum tunneling time of a bose-einstein condensate traversing
  through a laser-induced potential barrier.
\newblock {\em Phys. Rev. A}, 81:055602, May 2010.

\bibitem{BrightSolitonsTunneling}
L.~Salasnich, A.~Parola, and L.~Reatto.
\newblock Condensate bright solitons under transverse confinement.
\newblock {\em Phys. Rev. A}, 66:043603, Oct 2002.

\bibitem{SolitonInterferometerTunneling2013}
J.~Polo and V.~Ahufinger.
\newblock Soliton-based matter-wave interferometer.
\newblock {\em Phys. Rev. A}, 88:053628, Nov 2013.

\bibitem{marchant_quantum_2016}
A.~L. Marchant, T.~P. Billam, M.~M.~H. Yu, A.~Rakonjac, J.~L. Helm, J.~Polo,
  C.~Weiss, S.~A. Gardiner, and S.~L. Cornish.
\newblock Quantum reflection of bright solitary matter waves from a narrow
  attractive potential.
\newblock {\em Physical Review A}, 93(2):021604, February 2016.

\bibitem{BECPulsedTunneling}
L.~Salasnich, A.~Parola, and L.~Reatto.
\newblock Pulsed macroscopic quantum tunneling of falling bose-einstein
  condensates.
\newblock {\em Phys. Rev. A}, 64:023601, Jun 2001.

\bibitem{BECInteractionAssistedTunnelingExperimental}
Shreyas Potnis, Ramon Ramos, Kenji Maeda, Lincoln~D. Carr, and Aephraim~M.
  Steinberg.
\newblock Interaction-assisted quantum tunneling of a bose-einstein condensate
  out of a single trapping well.
\newblock {\em Phys. Rev. Lett.}, 118:060402, Feb 2017.

\bibitem{GuidedAtomLaserAspect2007}
{Billy, J.}, {Josse, V.}, {Zuo, Z.}, {Guerin, W.}, {Aspect, A.}, and {Bouyer,
  P.}
\newblock Guided atom laser: a new tool for guided atom optics.
\newblock {\em Ann. Phys. Fr.}, 32(2-3):17--24, 2007.

\bibitem{kuhn_bose-condensed_2014}
C.~C.~N. Kuhn, G.~D. McDonald, K.~S. Hardman, S.~Bennetts, P.~J. Everitt, P.~A.
  Altin, J.~E. Debs, J.~D. Close, and N.~P. Robins.
\newblock A {Bose}-condensed, simultaneous dual-species {Mach}--{Zehnder} atom
  interferometer.
\newblock {\em New Journal of Physics}, 16(7):073035, 2014.

\bibitem{MI_Patrick_2017}
P.~J. Everitt, M.~A. Sooriyabandara, M.~Guasoni, P.~B. Wigley, C.~H. Wei, G.~D.
  McDonald, K.~S. Hardman, P.~Manju, J.~D. Close, C.~C.~N. Kuhn, S.~S. Szigeti,
  Y.~S. Kivshar, and N.~P. Robins.
\newblock Observation of a modulational instability in bose-einstein
  condensates.
\newblock {\em Phys. Rev. A}, 96:041601, Oct 2017.

\bibitem{Imaging_Paul2016}
P.~B. Wigley, P.~J. Everitt, K.~S. Hardman, M.~R. Hush, C.~H. Wei, M.~A.
  Sooriyabandara, P.~Manju, J.~D. Close, N.~P. Robins, and C.~C.~N. Kuhn.
\newblock Non-destructive shadowgraph imaging of ultra-cold atoms.
\newblock {\em Opt. Lett.}, 41(20):4795--4798, Oct 2016.

\bibitem{SolitonGordonPRL2014}
G.~D. McDonald, C.~C.~N. Kuhn, K.~S. Hardman, S.~Bennetts, P.~J. Everitt, P.~A.
  Altin, J.~E. Debs, J.~D. Close, and N.~P. Robins.
\newblock Bright solitonic matter-wave interferometer.
\newblock {\em Phys. Rev. Lett.}, 113:013002, Jul 2014.

\bibitem{ErratumSolitonGordonPRL}
G.~D. McDonald, C.~C.~N. Kuhn, K.~S. Hardman, S.~Bennetts, P.~J. Everitt, P.~A.
  Altin, J.~E. Debs, J.~D. Close, and N.~P. Robins.
\newblock Erratum: Bright solitonic matter-wave interferometer [phys. rev.
  lett. 113, 013002 (2014)].
\newblock {\em Phys. Rev. Lett.}, 118:219903, May 2017.

\bibitem{Ernst:2009}
Philipp~T. Ernst, S{\"o}ren G{\"o}tze, Jasper~S. Krauser, Karsten Pyka,
  Dirk-S{\"o}ren L{\"u}hmann, Daniela Pfannkuche, and Klaus Sengstock.
\newblock Probing superfluids in optical lattices by momentum-resolved bragg
  spectroscopy.
\newblock {\em Nature Physics}, 6:56 EP --, 11 2009.

\bibitem{Altin:2013}
P~A Altin, M~T Johnsson, V~Negnevitsky, G~R Dennis, R~P Anderson, J~E Debs, S~S
  Szigeti, K~S Hardman, S~Bennetts, G~D McDonald, L~D Turner, J~D Close, and
  N~P Robins.
\newblock Precision atomic gravimeter based on bragg diffraction.
\newblock {\em New Journal of Physics}, 15(2):023009, 2013.

\bibitem{Effective1DGPEPRL98}
M.~Olshanii.
\newblock Atomic scattering in the presence of an external confinement and a
  gas of impenetrable bosons.
\newblock {\em Phys. Rev. Lett.}, 81:938--941, Aug 1998.

\bibitem{Effective1DGPEPRA2002}
L.~Salasnich, A.~Parola, and L.~Reatto.
\newblock Effective wave equations for the dynamics of cigar-shaped and
  disk-shaped bose condensates.
\newblock {\em Phys. Rev. A}, 65:043614, Apr 2002.

\bibitem{DimensionalDeduction2015}
L.Salasnich and S.K.Adhikari.
\newblock Dimensional reduction and localization of a bose--einsteincondensate
  in a quasi-1d bichromatic optical lattice.
\newblock {\em Acta Physica Polonica A}, 128(6):979--982, May 2015.

\bibitem{ComputeBECGroundStateImaginaryTime}
W.~Bao and Q.~Du.
\newblock Computing the ground state solution of bose--einstein condensates by
  a normalized gradient flow.
\newblock {\em SIAM Journal on Scientific Computing}, 25(5):1674--1697, 2004.

\bibitem{SolveScrodingerEqImaginaryTime}
L.~Lehtovaara, J.~Toivanen, and J.~Eloranta.
\newblock Solution of time-independent schr{\"o}dinger equation by the
  imaginary time propagation method.
\newblock {\em Journal of Computational Physics}, 221(1):148 -- 157, 2007.

\bibitem{MeanFieldLimitVlasovBoltzmann}
Fran{\c c}ois Golse.
\newblock The mean-field limit for the dynamics of large particle systems.
\newblock {\em Journ{\'e}es {\'e}quations aux d{\'e}riv{\'e}es partielles},
  pages 1--47, 2003.

\bibitem{MeanFieldWithCollisionPRA}
P.~Pedri, D.~Gu\'ery-Odelin, and S.~Stringari.
\newblock Dynamics of a classical gas including dissipative and mean-field
  effects.
\newblock {\em Phys. Rev. A}, 68:043608, Oct 2003.

\bibitem{TransportPropertiesLhuillierLaloe1982}
C.~Lhuillier et~F.~Lalo{\"e}.
\newblock Transport properties in a spin polarized gas, i.
\newblock {\em Journal de Physique}, 43(Feburary):197--224, 1982.

\bibitem{KernelEstimationPhysics}
Kyle Cranmer.
\newblock Kernel estimation in high-energy physics☆☆this work was partially
  supported by a graduate research fellowship from the national science
  foundation and us department of energy grant de-fg0295-er40896.
\newblock {\em Computer Physics Communications}, 136(3):198 -- 207, 2001.

\bibitem{Mocz:2015}
Philip Mocz and Sauro Succi.
\newblock Numerical solution of the nonlinear schr\"odinger equation using
  smoothed-particle hydrodynamics.
\newblock {\em Phys. Rev. E}, 91:053304, May 2015.

\bibitem{Jiang:2018}
Tao Jiang, Zhen-Chao Chen, Wei-Gang Lu, Jin-Yun Yuan, and Deng-Shan Wang.
\newblock An efficient split-step and implicit pure mesh-free method for the
  2d/3d nonlinear gross--pitaevskii equations.
\newblock {\em Computer Physics Communications}, 231:19--30, 2018.

\bibitem{Guery-Odelin:2002}
D.~Gu\'ery-Odelin.
\newblock Mean-field effects in a trapped gas.
\newblock {\em Phys. Rev. A}, 66:033613, Sep 2002.

\bibitem{Bouchoule:2016}
I.~Bouchoule, S.~S. Szigeti, M.~J. Davis, and K.~V. Kheruntsyan.
\newblock Finite-temperature hydrodynamics for one-dimensional bose gases:
  Breathing-mode oscillations as a case study.
\newblock {\em Phys. Rev. A}, 94:051602, Nov 2016.

\bibitem{BrightSolitonTunnelingAttractiveBarrierDOdelin2016}
F.~Damon, B.~Georgeot, and D.~Gu{\'e}ry-Odelin.
\newblock Probing surface states with many-body wave packet scattering.
\newblock {\em EPL (Europhysics Letters)}, 115(2):20010, 2016.

\bibitem{BECNonlinearDynamicsTunneling}
G.~Dekel, O.V. Farberovich, A.~Soffer, and V.~Fleurov.
\newblock Nonlinear dynamic phenomena in macroscopic tunneling.
\newblock {\em Physica D: Nonlinear Phenomena}, 238(15):1475 -- 1481, 2009.
\newblock Nonlinear Phenomena in Degenerate Quantum Gases.

\bibitem{gauthier_direct_2016}
G.~Gauthier, I.~Lenton, N.~McKay Parry, M.~Baker, M.~J. Davis,
  H.~Rubinsztein-Dunlop, and T.~W. Neely.
\newblock Direct imaging of a digital-micromirror device for configurable
  microscopic optical potentials.
\newblock {\em Optica}, 3(10):1136--1143, October 2016.

\bibitem{SuperluminalTraversalTime}
Xi~Chen and Chun-Fang Li.
\newblock Superluminal traversal time and interference between multiple finite
  wave packets.
\newblock {\em EPL (Europhysics Letters)}, 82(3):30009, 2008.

\bibitem{Damon:2014}
F.~Damon, F.~Vermersch, J.~G. Muga, and D.~Gu\'ery-Odelin.
\newblock Reduction of local velocity spreads by linear potentials.
\newblock {\em Phys. Rev. A}, 89:053626, May 2014.

\bibitem{SpaceMomRepresentationHartmann}
Mar\'{\i}a~F. Gonz\'alez, Josep~Maria Bofill, Xavier Gim\'enez, and F.~Borondo.
\newblock Space and momentum representation analysis of hartman's effect in
  wave packet transmission.
\newblock {\em Phys. Rev. A}, 78:032102, Sep 2008.

\end{thebibliography}

\end{document}